\pdfoutput=1

\documentclass[11pt,twoside,a4paper,cmspaper,final,collab]{cms-tdr}
\def\svnVersion{168105}\def\cmsCernNo{2012-235}\def\cmsCernDate{\today}\def\cmsMessage{Submitted to Physical Review Letters}

\begin{document}\cmsNoteHeader{EXO-11-091}

\hyphenation{had-ron-i-za-tion}
\hyphenation{cal-or-i-me-ter}
\hyphenation{de-vices}

\RCS$Revision: 158350 $
\RCS$HeadURL: svn+ssh://svn.cern.ch/reps/tdr2/papers/EXO-11-091/trunk/EXO-11-091.tex $
\RCS$Id: EXO-11-091.tex 158350 2012-11-18 20:00:02Z alverson $
\ifthenelse{\boolean{cms@external}}{\providecommand{\suppMaterial}{the supplemental material [URL will be inserted by publisher]}}{\providecommand{\suppMaterial}{App.~\ref{app:suppMat}}}
\cmsNoteHeader{EXO-11-091} 
\newcommand{\WR}{\ensuremath{\PW_{\cmsSymbolFace{R}}}\xspace}
\newcommand{\ZR}{\ensuremath{\cPZ_{\cmsSymbolFace{R}}}\xspace}
\newcommand{\Nmu}{\ensuremath{\cmsSymbolFace{N}_{\Pgm}}\xspace}
\newcommand{\Nell}{\ensuremath{\cmsSymbolFace{N}_{\ell}}\xspace}
\newcommand{\Ne}{\ensuremath{\cmsSymbolFace{N}_{\Pe}}\xspace}
\newcommand{\re}{\ensuremath{\cmsSymbolFace{e}}}
\newcommand{\KL}{\ensuremath{\PK_{\cmsSymbolFace{L}}}\xspace}
\newcommand{\KS}{\ensuremath{\PK_{\cmsSymbolFace{S}}}\xspace}
\providecommand{\POWHEG}{{\textsc{powheg}}\xspace}
\title{Search for heavy neutrinos and \texorpdfstring{\WR}{W[R]} bosons with right-handed couplings in a left-right symmetric model in \texorpdfstring{$\Pp\Pp$ collisions at $\sqrt{s} = 7\TeV$}{pp collisions at sqrt(s) = 7 TeV}}

\date{\today}

\abstract{
Results are presented from a search for heavy, right-handed muon neutrinos, \Nmu, and right-handed \WR\ bosons, which arise in the left-right symmetric extensions of the standard model.  The analysis is based on a 5.0\fbinv sample of proton-proton collisions at a center-of-mass energy of 7\TeV, collected by the CMS detector at the Large Hadron Collider.  No evidence is observed for an excess of events over the standard model expectation.  For models with exact left-right symmetry, heavy right-handed neutrinos are excluded at 95\% confidence level for a range of neutrino masses below the $\WR$ mass, dependent on the value of $M_{\WR}$.  The excluded region in the two-dimensional $(M_{\WR}, M_{\Nmu})$ mass plane extends to $M_{\WR} = 2.5\TeV$.
}

\hypersetup{%
pdfauthor={CMS Collaboration},%
pdftitle={Search for heavy neutrinos and W bosons with right-handed couplings in a left-right symmetric model in pp collisions at 7 TeV},%
pdfsubject={CMS},%
pdfkeywords={CMS, physics}}

\maketitle 

The maximal violation of parity conservation is a
prominent feature of neutrino interactions that is included
in the standard model (SM) in terms of purely
left-handed couplings to the \PW\ boson.  In addition,
the observation of neutrino oscillations
(see e.g.~\cite{nu}), together with direct limits on
neutrino masses~\cite{numass}, has demonstrated that neutrinos
have tiny but non-vanishing masses, suggesting a distinct origin
from the masses of the quarks and leptons.

The left-right (LR) symmetric extension of the
standard model~\cite{lr,lr1,lr2,lr3} provides a
possible explanation for neutrino mass
through the see-saw mechanism~\cite{seesaw}. The LR symmetry
is spontaneously broken at a multi-TeV mass
scale, leading to parity violation in weak interactions
as described by the SM.  By introducing a right-handed
SU(2) symmetry group, the LR model incorporates
heavy right-handed Majorana neutrinos (\Nell, $\ell = \Pe, \mu, \tau$)
as well as additional charged ($\WR^{\pm}$)
and neutral (\ZR) gauge bosons.

We search for the production of \WR bosons from proton-proton
collisions at the Large Hadron Collider (LHC).
The \WR boson is assumed to decay to a muon and to a
right-handed neutrino \Nmu, which subsequently
decays to produce a second muon together with a virtual $\WR^*$.
If the \Nmu is a Majorana particle as predicted in the LR model,
the two final state muons may have the same sign.
The virtual $\WR^*$ decays to a pair of quarks which
hadronize into jets ($j$), resulting in
a final state with two muons and two jets:

\begin{equation*}
\WR \rightarrow \mu_1 \Nmu \rightarrow \mu_1 \mu_2 \WR^* \rightarrow \mu_1 \mu_2 \cPq \cPq' \rightarrow \mu_1 \mu_2 j_1 j_2.
\end{equation*}

The search presented in this Letter
is characterized by the \WR and \Nmu masses,
$M_{\WR}$ and $M_{\Nmu}$,
which are allowed to vary independently.  Although
$M_{\Nmu} > M_{\WR}$ is allowed, it is not considered
in this analysis.  The
branching fraction for $\WR \rightarrow \mu \Nmu$ depends on the
number of heavy neutrino flavors that are accessible at
LHC energies.
To simplify the interpretation of the
results, \Nmu is assumed to be
the only heavy neutrino flavor
light enough to contribute significantly to the \WR decay width.
CMS recently performed a search for heavy Majorana
neutrinos in the final state containing two jets and
two same-sign electrons or muons and set limits on the coupling
between such a neutrino and the left-handed \PW\ of the SM
as a function of $M_{\Nmu}$~\cite{EXO-11-076}, while this
analysis considers on-shell production of a
right-handed \WR boson.  No charge requirements
are imposed on the final state muons in this analysis.

For given \WR and \Nmu masses,
the signal cross section can be predicted from the
assumed value of the coupling constant $g_R$,
which denotes the strength of the gauge interactions
of $\WR^{\pm}$ bosons.  Strict left-right symmetry implies
that $g_R$ is equal to the (left-handed) weak
interaction coupling strength $g_L$ at $M_{\WR}$,
which will be assumed throughout this Letter.
Consequently, the \WR production cross section can be
calculated by
the \textsc{fewz} program~\cite{kfactor} using the
left-handed $\PWpr$ model~\cite{wpr1,wpr2}.
As an additional simplification,
the left-right boson
and lepton
mixing angles are assumed to be small.

Estimates based on $\KL - \KS$
mixing results imply a
theoretical lower limit of
$M_{\WR} \gtrsim 2.5$\TeV~\cite{mix1,mix2}.
Searches for $\WR \rightarrow \cPqt\cPqb$ decays at the
Tevatron~\cite{d0,cdf,d0new}
and at the LHC~\cite{atlastb,cmstb}
exclude \WR masses below 1.85\TeV.
An ATLAS search for
$\WR \rightarrow \ell \Nell$
using similar model assumptions as those in this Letter,
but allowing \WR decays to both \Ne and \Nmu,
excluded a region in the two-dimensional
parameter $(M_{\WR},M_{\Nell})$ space
extending to nearly $M_{\WR} = 2.5$\TeV~\cite{atlas}.

The analysis is based on a 5.0\fbinv sample of
proton-proton collision data at a center-of-mass energy
of 7\TeV, collected by the Compact Muon Solenoid (CMS)
detector~\cite{CMS} at the LHC.
The central feature of the CMS apparatus is a
superconducting solenoid, of 6~m internal diameter, providing a field
of 3.8\unit{T}. Within the field volume are the silicon pixel and strip
trackers, the lead-tungstate crystal electromagnetic calorimeter,
and the brass/scintillator hadron calorimeter.
Muons are measured in gas-ionization detectors embedded in the steel return
yoke, with detection planes made of
three technologies: drift tubes, cathode strip chambers, and resistive
plate chambers. 
The CMS trigger system, composed of custom hardware processors
at the first level followed by a processor farm at the next level,
selects $\mathcal{O}(100\unit{Hz})$ of the most interesting
events.
The events used in this analysis were collected with single-muon
triggers whose \pt thresholds ranged from 24\GeV to 40\GeV, depending
on the instantaneous luminosity.

The $\WR \rightarrow \mu \Nmu$ signal samples are generated using
\PYTHIA 6.4.24~\cite{pythia}, which includes the LR symmetric
model with the standard assumptions mentioned previously,
with CTEQ6L1 parton distribution functions~\cite{parton}.
We also study SM background processes
using simulated samples:
\ttbar and single-top (both generated using \POWHEG~\cite{powheg}),
\PW\ and Drell--Yan production in association with jets
(\SHERPA~\cite{sherpa}), and diboson production (\PYTHIA).
Generated events pass through the full CMS detector
simulation based on \GEANTfour~\cite{geant4}.

The muon identification strategy is based on both the muon detectors
and the inner tracker, described in Ref.~\cite{CMS-PAS-MUO-10-002}.
At least one of the two muons used to define the \WR candidate is
required to be matched to a muon candidate found by the trigger,
and both muons are required to satisfy the
tight identification criteria discussed in Ref.~\cite{Zprime}.
The muon identification requirements ensure good consistency between
the measurements of the muon detector and the
inner tracker, and suppress muons from decay-in-flight of
hadrons as well as from shower punch-through.  Non-isolated muon
backgrounds are controlled by computing the sum of
the transverse momentum of tracks within a cone
about the muon direction of $\Delta R < 0.3$,
with $\Delta R = \sqrt{(\Delta \eta)^{2} + (\Delta \phi)^{2}}$,
given the azimuthal angle $\phi$ and
$\eta = -\ln [\tan(\theta/2)]$, where $\theta$ is the
polar angle with respect to the beam direction.
The final \pt sum must be
less than 10\% of the muon transverse momentum.

Jets are reconstructed by forming clusters of charged and neutral hadrons,
photons, and leptons that are first reconstructed based on
the CMS particle-flow technique~\cite{pf},
using the anti-\kt clustering algorithm~\cite{antikt} with a
radius parameter $R=0.5$. 
Energy deposits in the calorimeter with characteristics
that match those of noise or beam halo tracks are identified,
and events are rejected
if either of the two highest-\pt jet
candidates was produced by such energy deposits.
To suppress backgrounds from heavy-flavor-quark decays,
any muon is rejected if found near a jet,
with $\Delta R(\mu,j) < 0.5$.

In approximately 95\% of simulated signal event samples, the \WR
final state decay products are the highest \pt muons
and jets in the event.
$\WR \rightarrow \mu \Nmu$ candidates are thus
formed from the two highest-\pt
muons and the two highest-\pt jets in the event.
As the initial two-body decay
$\WR \rightarrow \mu \Nmu$ tends to produce a
high-momentum muon, events are selected in which the
leading muon has $\pt > 60$\GeV and the subleading muon
has $\pt > 30$\GeV.
A minimum transverse momentum requirement of 40\GeV is
imposed on the jet
candidates after correcting for the effects of the
extra pp collisions in the event
and the jet energy response of the detector.
Backgrounds are suppressed by
requiring the invariant mass of the dimuon system
$M_{\mu \mu} > 200$\GeV
and the four-object mass $M_{\mu \mu jj} > 600$\GeV.

The signal acceptance is found to be typically near 80\% at
$M_{\Nmu} \sim M_{\WR}/2$ and decreases rapidly for $M_{\Nmu} \lesssim 0.10 M_{\WR}$.
At low neutrino mass, the $\Nmu \rightarrow \mu jj$
decay products tend to overlap due to the boost
from \WR decay, and the two jets may not be distinguishable or the
muon from \Nmu decay may be too close to a jet.
For \WR signal events which meet the kinematic
acceptance requirements, the efficiency to reconstruct
the four high-\pt objects using the CMS
detector ranges between 75\% and 80\% as a function
of \WR and \Nmu mass.

After the muon requirements are applied, the SM backgrounds for
$\WR \rightarrow \mu \Nmu$ consist primarily of
events from processes with two isolated high-\pt muons, namely
$\ttbar \rightarrow \cPqb \PWp \cPaqb \PWm$ and \cPZ+jets
processes.  The impact of the selection criteria on background
processes is shown in Table~\ref{tab:numbers_select}.

The \ttbar background contribution is estimated using a
control sample of $\Pe \mu jj$ events reconstructed in data
and simulation.  This sample is dominated by \ttbar events,
with small contributions from other SM processes estimated
using simulation.  The simulated \ttbar background
contribution is scaled to data using events
satisfying $M_{\Pe \mu} > 200$\GeV, which is equivalent to the
third selection stage in Table~\ref{tab:numbers_select}.
The scale factor for the simulated \ttbar sample, relative
to the \ttbar cross section measured by CMS~\cite{topCS},
is $0.97 \pm 0.06$.  The uncertainty on this scale
factor reflects the number of events in data
with $M_{\Pe \mu} > 200$\GeV. Applying this scale factor to the
\ttbar simulation, the $M_{\Pe \mu j j}$ distributions
in data and simulation are found to be in agreement.
This scale factor is applied to the simulated $\ttbar$
event sample at all stages of selection in order to estimate
the expected number of $\Pp\Pp \rightarrow \ttbar + \mathrm{X}$
events that survive successive selection criteria.

The \cPZ+jets background contribution is estimated from
$\cPZ\rightarrow \mu\mu$ decays reconstructed
in simulation and data.  The simulated \cPZ+jets background
contribution is normalized
to data using events in the dimuon mass region
$60\GeV < M_{\mu \mu} < 120\GeV$ after requiring $\mu_1$
$\pt > 60$\GeV as indicated in Table~\ref{tab:numbers_select}.
Accounting for other SM background processes, the simulated
\cPZ+jets scale factor is $1.43 \pm 0.01$ relative to
inclusive next-to-next-to-leading order calculations.  The
uncertainty on this value reflects the number of
events from data with $60\GeV < M_{\mu \mu} < 120\GeV$.
After rescaling the \cPZ+jets simulation, the shape
of the $M_{\mu \mu}$ distribution for data is in
agreement with simulation for $M_{\mu \mu} > 60$\GeV.

After all selection criteria are applied, the \ttbar and \cPZ+jets
processes dominate the total SM background contribution.  Other
SM processes, mostly diboson and single top, comprise less
than 5\% of the total background and their contributions are
estimated from simulation.  Background from \PW+jets processes,
also estimated from simulation, is negligible.
The background contribution from multijet
processes is estimated using control samples from data and is
roughly 0.1\% of the total SM background after all
selection requirements are applied.

The observed and expected number of events surviving
the selections are summarized in Table~\ref{tab:numbers_select}.
The yields
reflect the number of background events surviving each
selection stage, with normalization factors obtained from
control sample studies (\ttbar, \cPZ+jets, and multijet processes)
or taken directly from simulation.
The data are found to be in agreement with SM expectations.

\begin{table*}
\begin{center}
  \topcaption{The total number of events reconstructed in data,
  and the expected contributions
  from signal and background (bkgd) samples,
  after different stages of the selection
    requirements are applied.  The first selection given below requires two muons with $\pt > 30$\GeV
    and two jets with $\pt > 40$\GeV meeting all requirements described in the text.
    The ``Signal"
    column indicates the expected contribution
    for $M_{\WR} = 1800$\GeV, with $M_{\Nmu} = 1000$\GeV.
    The uncertainties for the background
    expectation are derived for the final stage of selection and
    more details are given in the text.
    The yields from earlier stages of the selection
    have greater relative uncertainty than that for the
    full selection.
}

\begin{tabular}{lcccccc} \hline
Selection stage         &  Data  & Signal &   Total bkgd   & \ttbar      & \cPZ+jets      & Other \\ \hline
Two muons, two jets       & 21769  &  50            &   21061      &   1603      &    19136    & 322  \\ 
$\mu_1$ $\pt > 60$\GeV   & 13328  &  50            &   12862      &   1106      &    11531    & 225  \\ 
$M_{\mu\mu} > 200$\GeV    &   365  &  48           &     341      &    211      &      116    & 14  \\ 
$M_{\mu\mu jj} > 600$\GeV &   164  &  $48 \pm 13$  & $152 \pm 22$   & $81 \pm 18$ & $65 \pm 9$ & $6 \pm 3$  \\ \hline
\end{tabular}
\label{tab:numbers_select}
\end{center}
\end{table*}

The reconstructed four-object mass in data and simulation
is used to estimate limits on \WR production.
The $M_{\mu \mu jj}$ distribution for
$\WR \rightarrow \mu \mu j j$ signal events, for each
\WR mass assumption, is included together with
the SM background distributions to search for evidence of
\WR production.

The dominant uncertainty related to
$\WR \rightarrow \mu \Nmu$ production arises from the variation
in the predicted signal production cross section
as a result of the uncertainties in the parton distribution
functions (PDFs) of the proton.  This uncertainty
varies between 4\% and 22\%, depending on
the \WR mass hypothesis, following the
PDF4LHC prescriptions~\cite{pdf4lhc} for the
CT10~\cite{ct10} and MSTW2008~\cite{mstw} PDF sets.

The uncertainties associated with muon reconstruction
and identification
are determined from $\cPZ \rightarrow \Pgmp\Pgmm$ events
reconstructed in both data and simulation.
The size of this uncertainty is about 15\% for
signal and 5\% for background processes.

The shape of each SM background $M_{\mu \mu j j}$ distribution
is modeled by an exponential ($\re^{a+bM_{\mu \mu jj}}$) lineshape,
and the background contributions as a function of mass are
determined from the result of fits applied to each
background type: \ttbar, \cPZ+jets, and other SM backgrounds.
The background uncertainty is dominated
by the uncertainty in the background modeling
and is computed as a function of $\mu \mu jj$ mass.

The uncertainty in the exponential fit is taken as the
uncertainty due to background modeling.
Each background distribution is also fit
with an alternative suite of exponential functions to
allow for deviations from the assumed
shape at high mass.
For a given $M_{\mu \mu j j}$ range, we take the maximum
of the deviation, relative to the nominal exponential fit,
from any alternative fit result as the uncertainty due
to background modeling if this deviation exceeds the
nominal fit uncertainty.

Uncertainties in the jet energy scale and resolution
impact the shape of the signal and background
$M_{\mu \mu jj}$ distributions, contributing less than 10\%
to the signal and background uncertainties.
The normalization of the various background samples contributes
5\% to the total uncertainty.
Muon resolution and trigger efficiency uncertainties, and
additional factorization and scale theoretical uncertainties,
contribute to the total uncertainty to a lesser extent.
The uncertainties in the total number of background
events are derived taking into account the relative
contribution of all background events after the full event
selection, and the correlation of each
effect between all background processes.

\begin{figure}[!htb]
\includegraphics[width=0.90\linewidth]{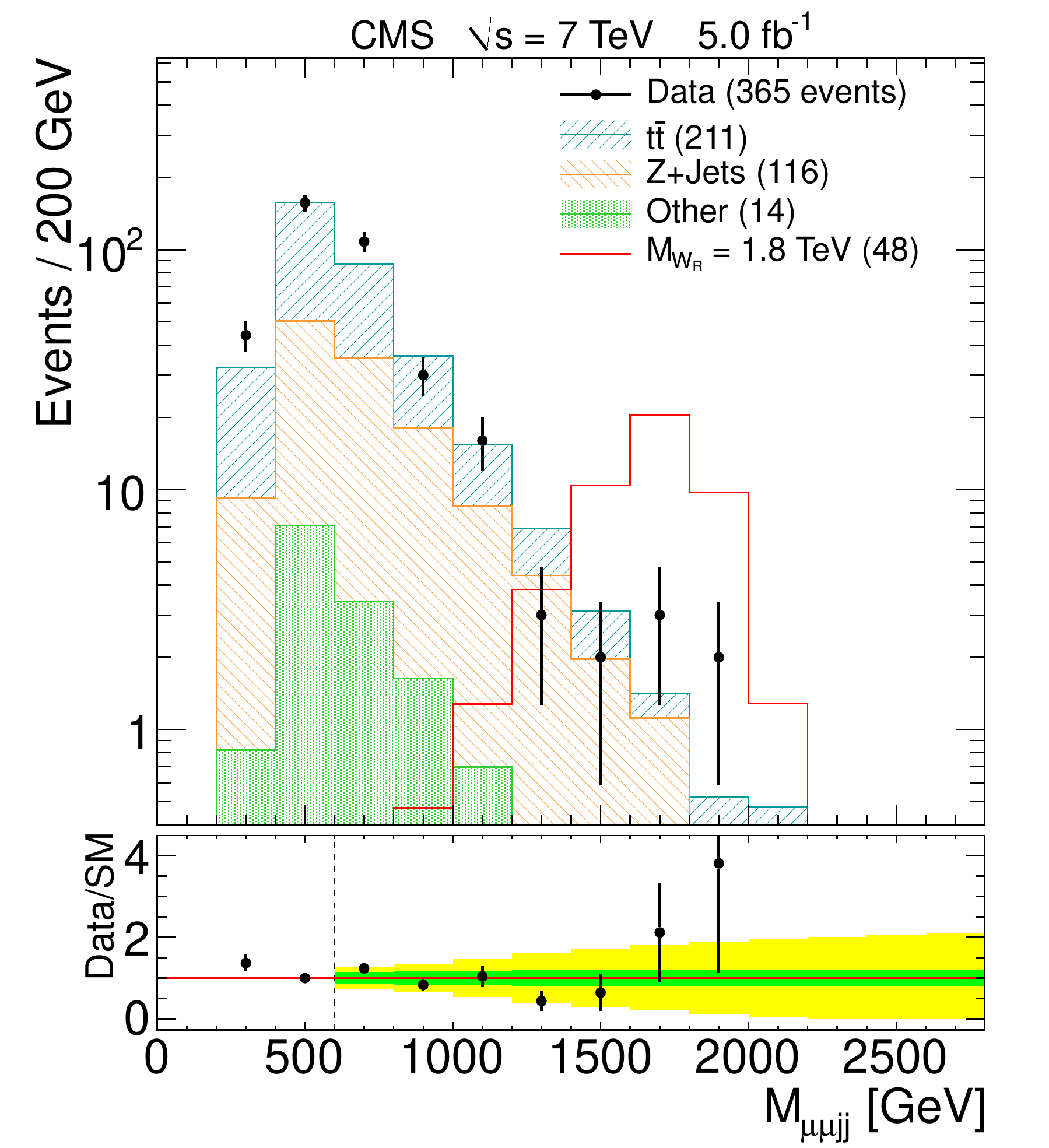}
\caption{Distribution of the invariant mass $M_{\mu \mu j j}$
for events in data (points with error bars)
with $M_{\mu \mu} > 200$\GeV
and for simulated background
contributions (hatched stacked histograms).
The signal mass point
$M_{\WR}=1800$\GeV, $M_{\Nmu}=1000$\GeV,
is included for comparison (open red histogram).
The number of events from each background process
(and the expected number of signal events) is
included in parentheses in the legend.
The data are compared to SM
expectations in the lower portion of the figure.  The total
background uncertainty (yellow band)
and the background uncertainty after neglecting
the uncertainty due to background modeling
(green band) are included as a function of
$M_{\mu \mu j j}$ for $M_{\mu \mu j j} > 600$\GeV.
\label{fig:evtsPostMll}
}
\end{figure}

The total uncertainty for signal and background is
summarized in Table~\ref{tab:numbers_select}.
The $M_{\mu \mu j j}$ distribution for events
with $M_{\mu \mu} > 200$\GeV is presented
in Fig.~\ref{fig:evtsPostMll}, which also
summarizes the background uncertainty as a function of
$M_{\mu \mu jj}$ and demonstrates the dominant background model
uncertainty relative to the total background uncertainty.

As no evidence for $\WR \rightarrow \mu \Nmu$ decay is found,
limits on \WR production are estimated using a multibin
technique based on the RooStats package~\cite{Moneta:2010pm}.
The bin width of 200\GeV, comparable to the mass resolution for a
reconstructed \WR boson with mass below 2.5\TeV,
is chosen for the $M_{\mu \mu j j}$ distributions
used to compute the limits.  The background inputs
to the limit calculation use the results of the exponential fit,
while the signal input is taken directly from
the $M_{\mu \mu j j}$ distribution for each signal \WR mass
assumption.  Uncertainties are included as nuisance parameters
in the limit calculations.  A $\mathrm{CL}_\mathrm{S}$
limit setting technique~\cite{cls1,cls2} is used to estimate the
95\% confidence level (CL) excluded region as a function of
the \WR cross section multiplied
by the $\WR \rightarrow \mu \mu jj$ branching fraction
and \WR mass.  The observed and expected limits are found
to be in agreement.  These
results (available in tabular form in \suppMaterial)
can be used for the evaluation of models
other than those considered in this Letter.

\begin{figure}[!htb]
\includegraphics[width=0.90\linewidth]{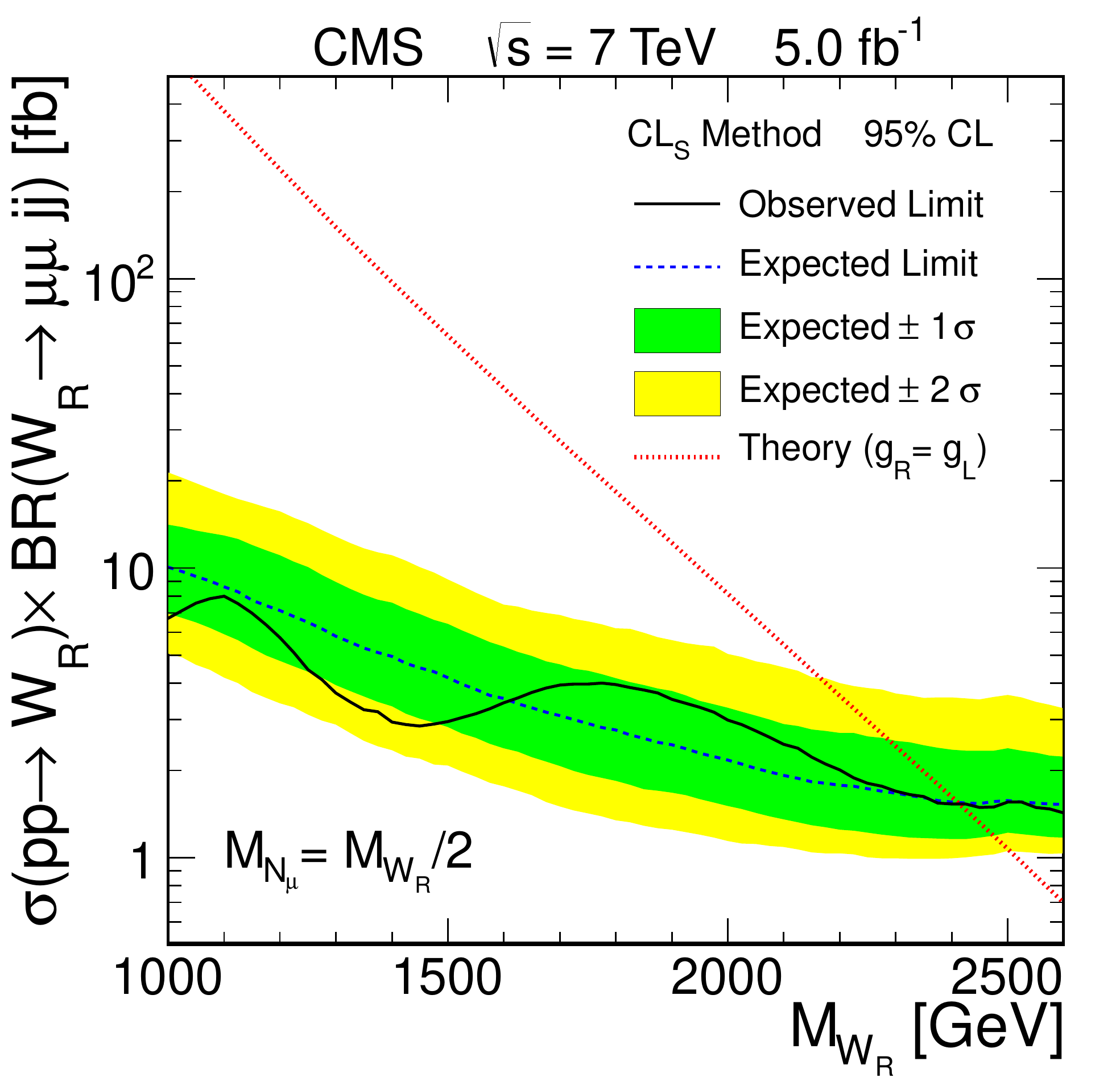}
\caption{The 95\% confidence level exclusion
limit on the \WR production cross section times
branching fraction for $\WR \rightarrow \mu \mu j j$
as a function of $M_{\WR}$ for
$M_{\Nmu} = \frac{1}{2} M_{\WR}$.  This limit is compared to
expectations given the theoretical model described
in the text.}
\label{fig:masslimitsWR}
\end{figure}

Limits as a function of \WR mass for
a right-handed neutrino with $M_{\cmsSymbolFace{N}} = \frac{1}{2} M_{\WR}$
are presented in Fig.~\ref{fig:masslimitsWR}.
The theoretical expectation in Fig.~\ref{fig:masslimitsWR}
assumes that only \Nmu contributes to the \WR decay width,
as mentioned previously.  Assuming degenerate $\Nell$
($\ell = \Pe, \mu, \tau$) masses allows
$\WR \rightarrow \Pe \Ne$ and
$\WR \rightarrow \tau \cmsSymbolFace{N}_{\tau}$
decays in addition to $\WR \rightarrow \cPq \cPaq'$ and
$\WR \rightarrow \mu \Nmu$ and effectively decreases
the expected $\WR \rightarrow \mu \mu j j$
production rate by approximately 15\%.

\begin{figure}[!htb]
\includegraphics[width=0.90\linewidth]{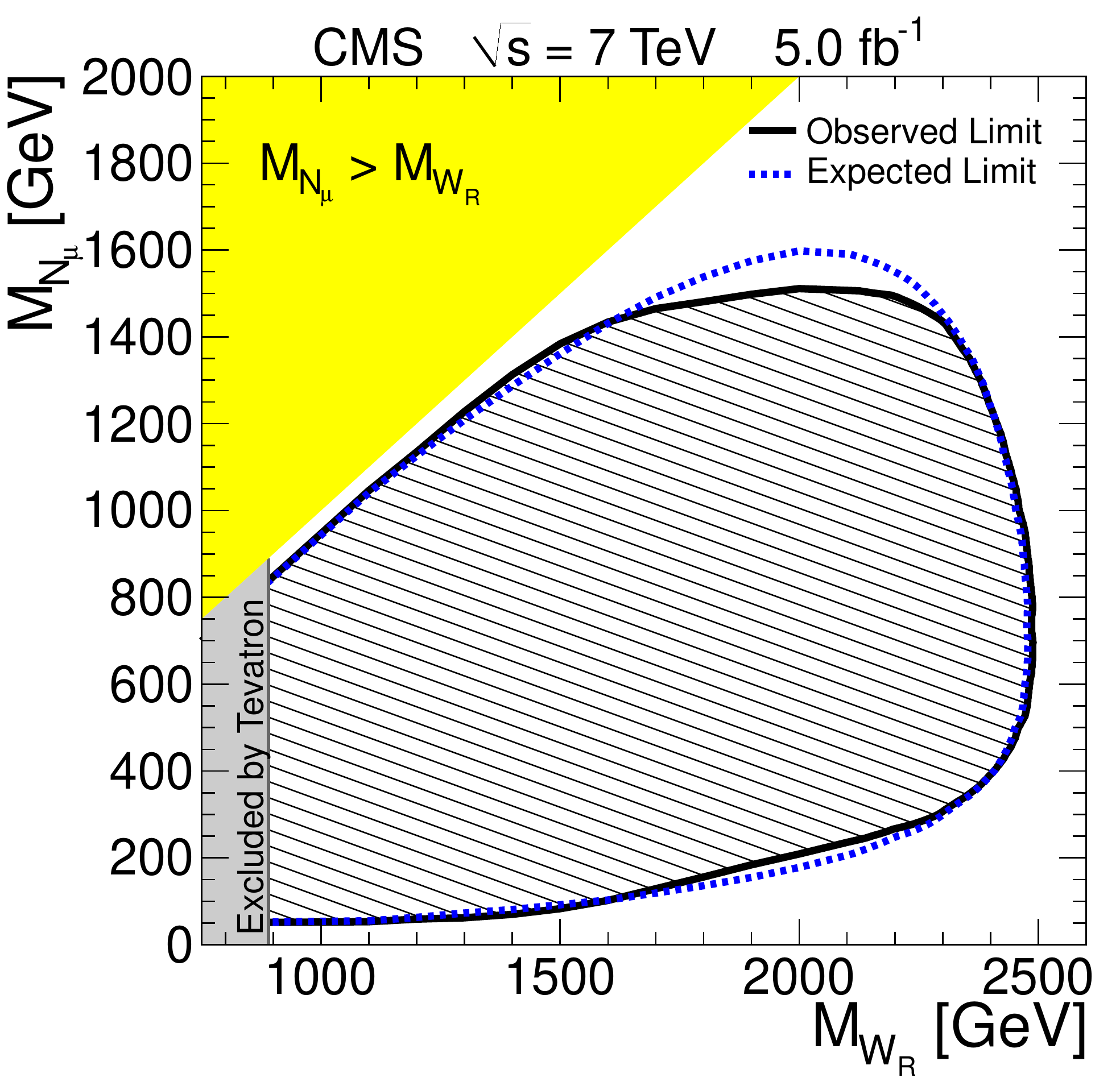}
\caption{The 95\% confidence level exclusion region in the
$(M_{\WR}, M_{\Nmu})$ plane, assuming the model
described in the text.  The Tevatron exclusion region for
\WR production~\cite{d0new} is included in the figure.}
\label{fig:masslimits2D}
\end{figure}

For the model considered in this Letter,
Fig.~\ref{fig:masslimits2D} indicates the range of
excluded \Nmu masses as a function of \WR mass
by comparing the observed (expected) upper limit and the predicted
cross section for each mass point.  These limits extend
to $M_{\WR} = 2.5$\TeV, and exclude a wide range of heavy
neutrino masses for \WR mass assumptions
below this maximal value.

In summary, we have presented a search for the
right-handed heavy muon neutrinos (\Nmu) and bosons (\WR)
of the left-right symmetric extension of the
standard model. We find that our
data sample is in agreement with expectations
from standard model
processes and therefore set a limit on the \WR and
\Nmu masses.  For models with exact
left-right symmetry (the same coupling to the right-handed
and left-handed sectors), we exclude heavy right-handed
neutrinos for a range of $M_{\Nmu} < M_{\WR}$, dependent
on the value of $M_{\WR}$.  For these models,
the excluded region in the two-dimensional parameter space
($M_{\WR}$,$M_{\Nmu}$) extends to $M_{\WR}= 2.5$\TeV.
These results represent the most sensitive limits to date on
\WR production assuming a single heavy neutrino flavor
contributes significantly to the \WR decay width.

We congratulate our colleagues in the CERN accelerator departments for the excellent performance of the LHC machine. We thank the technical and administrative staff at CERN and other CMS institutes, and acknowledge support from: BMWF and FWF (Austria); FNRS and FWO (Belgium); CNPq, CAPES, FAPERJ, and FAPESP (Brazil); MES (Bulgaria); CERN; CAS, MoST, and NSFC (China); COLCIENCIAS (Colombia); MSES (Croatia); RPF (Cyprus); MoER, SF0690030s09 and ERDF (Estonia); Academy of Finland, MEC, and HIP (Finland); CEA and CNRS/IN2P3 (France); BMBF, DFG, and HGF (Germany); GSRT (Greece); OTKA and NKTH (Hungary); DAE and DST (India); IPM (Iran); SFI (Ireland); INFN (Italy); NRF and WCU (Korea); LAS (Lithuania); CINVESTAV, CONACYT, SEP, and UASLP-FAI (Mexico); MSI (New Zealand); PAEC (Pakistan); MSHE and NSC (Poland); FCT (Portugal); JINR (Armenia, Belarus, Georgia, Ukraine, Uzbekistan); MON, RosAtom, RAS and RFBR (Russia); MSTD (Serbia); SEIDI and CPAN (Spain); Swiss Funding Agencies (Switzerland); NSC (Taipei); TUBITAK and TAEK (Turkey); STFC (United Kingdom); DOE and NSF (USA).

\bibliography{auto_generated}   
\ifthenelse{\boolean{cms@external}}{}{
\clearpage
\appendix
\section{95\% C.L. Exclusion Limits as a function of \texorpdfstring{\WR and \Nmu}{W[R] and N[R]} mass (tabular format)\label{app:suppMat}}

\begin{table}[!h]
\centering
\topcaption{The 95\% confidence level observed (Obs.) 
and expected (Exp.) exclusion limits (in fb) on the \WR 
production cross section times branching fraction for 
$\WR \rightarrow \mu \mu j j$ as a function of \WR and \Nmu 
mass (in~\GeV) for $800~\GeV \leq M_{\WR} \leq 1100~\GeV$.  
The 68\% and 95\% uncertainty bands for the expected limit 
(Exp. $\pm 1 \sigma$ and Exp. $\pm 2 \sigma$, respectively) 
are also included for each $(M_{\WR},M_{\Nmu})$ entry.} 
\begin{tabular}{cccccccc}
\hline
$M_{\WR}$   &  $M_{\Nmu}$  & Obs. Limit & Exp. Limit & Exp. $-1 \sigma$  & Exp. $+1 \sigma$ & Exp. $-2 \sigma$  & Exp. $+2 \sigma$  \\
\hline
800 & 100 & 164.05 & 117.93 & 83.22 & 174.17 & 59.49 & 238.37\\
800 & 200 & 48.16 & 34.62 & 24.43 & 51.13 & 17.47 & 69.98\\
800 & 300 & 34.80 & 25.02 & 17.66 & 36.95 & 12.62 & 50.57\\
800 & 400 & 30.14 & 21.67 & 15.29 & 32.00 & 10.93 & 43.80\\
800 & 500 & 28.54 & 20.52 & 14.48 & 30.30 & 10.35 & 41.47\\
800 & 600 & 29.96 & 21.54 & 15.20 & 31.81 & 10.86 & 43.53\\
800 & 700 & 40.88 & 29.39 & 20.74 & 43.41 & 14.83 & 59.41\\ \hline
900 & 100 & 51.43 & 83.52 & 58.94 & 114.27 & 43.95 & 143.72\\
900 & 200 & 13.52 & 21.96 & 15.49 & 30.04 & 11.55 & 37.78\\
900 & 300 & 9.50 & 15.43 & 10.89 & 21.11 & 8.12 & 26.55\\
900 & 400 & 8.00 & 12.98 & 9.16 & 17.77 & 6.83 & 22.34\\
900 & 500 & 7.35 & 11.93 & 8.42 & 16.32 & 6.28 & 20.53\\
900 & 600 & 7.28 & 11.83 & 8.35 & 16.19 & 6.23 & 20.36\\
900 & 700 & 7.81 & 12.68 & 8.95 & 17.35 & 6.67 & 21.82\\
900 & 800 & 10.19 & 16.54 & 11.67 & 22.63 & 8.70 & 28.46\\ \hline
1000 & 100 & 61.33 & 89.04 & 61.43 & 124.83 & 44.84 & 188.82\\
1000 & 200 & 13.85 & 20.11 & 13.88 & 28.20 & 10.13 & 42.65\\
1000 & 300 & 9.11 & 13.23 & 9.13 & 18.54 & 6.66 & 28.05\\
1000 & 400 & 7.62 & 11.07 & 7.64 & 15.52 & 5.57 & 23.47\\
1000 & 500 & 6.95 & 10.09 & 6.96 & 14.14 & 5.08 & 21.39\\
1000 & 600 & 6.67 & 9.68 & 6.68 & 13.58 & 4.88 & 20.54\\
1000 & 700 & 6.68 & 9.70 & 6.69 & 13.60 & 4.89 & 20.58\\
1000 & 800 & 7.19 & 10.44 & 7.20 & 14.63 & 5.26 & 22.14\\
1000 & 900 & 9.23 & 13.39 & 9.24 & 18.78 & 6.75 & 28.40\\ \hline
1100 & 100 & 75.54 & 90.83 & 62.35 & 136.58 & 44.07 & 190.23\\
1100 & 200 & 15.34 & 18.45 & 12.66 & 27.74 & 8.95 & 38.63\\
1100 & 300 & 10.11 & 12.16 & 8.34 & 18.28 & 5.90 & 25.46\\
1100 & 400 & 8.15 & 9.80 & 6.73 & 14.73 & 4.75 & 20.52\\
1100 & 500 & 7.36 & 8.85 & 6.07 & 13.30 & 4.29 & 18.53\\
1100 & 600 & 6.95 & 8.35 & 5.73 & 12.56 & 4.05 & 17.49\\
1100 & 700 & 6.82 & 8.20 & 5.63 & 12.34 & 3.98 & 17.18\\
1100 & 800 & 6.97 & 8.38 & 5.75 & 12.59 & 4.06 & 17.54\\
1100 & 900 & 7.38 & 8.88 & 6.10 & 13.35 & 4.31 & 18.60\\
1100 & 1000 & 9.18 & 11.04 & 7.58 & 16.60 & 5.36 & 23.12\\
\hline
\end{tabular}
\label{tab:limits800to1100}
\end{table}

\begin{table}[!h]
\centering
\topcaption{The 95\% confidence level observed (Obs.) 
and expected (Exp.) exclusion limits (in fb) on the \WR 
production cross section times branching fraction for 
$\WR \rightarrow \mu \mu j j$ as a function of \WR and \Nmu 
mass (in~\GeV) for $1200~\GeV \leq M_{\WR} \leq 1400~\GeV$.  
The 68\% and 95\% uncertainty bands for the expected limit 
(Exp. $\pm 1 \sigma$ and Exp. $\pm 2 \sigma$, respectively) 
are also included for each $(M_{\WR},M_{\Nmu})$ entry.} 
\begin{tabular}{cccccccc}
\hline
$M_{\WR}$   &  $M_{\Nmu}$  & Obs. Limit & Exp. Limit & Exp. $-1 \sigma$  & Exp. $+1 \sigma$ & Exp. $-2 \sigma$  & Exp. $+2 \sigma$  \\
\hline
1200 & 100 & 77.54 & 96.52 & 64.72 & 149.76 & 46.17 & 212.60\\
1200 & 200 & 14.31 & 17.81 & 11.94 & 27.63 & 8.52 & 39.22\\
1200 & 300 & 8.64 & 10.75 & 7.21 & 16.68 & 5.14 & 23.68\\
1200 & 400 & 6.92 & 8.62 & 5.78 & 13.37 & 4.12 & 18.98\\
1200 & 500 & 6.22 & 7.74 & 5.19 & 12.01 & 3.70 & 17.05\\
1200 & 600 & 5.73 & 7.13 & 4.78 & 11.06 & 3.41 & 15.70\\
1200 & 700 & 5.62 & 6.99 & 4.69 & 10.85 & 3.34 & 15.40\\
1200 & 800 & 5.57 & 6.94 & 4.65 & 10.76 & 3.32 & 15.28\\
1200 & 900 & 5.75 & 7.16 & 4.80 & 11.11 & 3.42 & 15.77\\
1200 & 1000 & 6.12 & 7.62 & 5.11 & 11.83 & 3.65 & 16.79\\
1200 & 1100 & 7.47 & 9.30 & 6.24 & 14.43 & 4.45 & 20.48\\ \hline
1300 & 100 & 56.46 & 83.58 & 56.80 & 128.85 & 41.46 & 184.69\\
1300 & 200 & 10.85 & 16.06 & 10.91 & 24.76 & 7.97 & 35.49\\
1300 & 300 & 6.16 & 9.12 & 6.20 & 14.06 & 4.52 & 20.15\\
1300 & 400 & 4.90 & 7.26 & 4.93 & 11.18 & 3.60 & 16.03\\
1300 & 500 & 4.31 & 6.39 & 4.34 & 9.84 & 3.17 & 14.11\\
1300 & 600 & 3.96 & 5.86 & 3.98 & 9.04 & 2.91 & 12.95\\
1300 & 700 & 3.86 & 5.71 & 3.88 & 8.80 & 2.83 & 12.61\\
1300 & 800 & 3.79 & 5.61 & 3.81 & 8.65 & 2.78 & 12.40\\
1300 & 900 & 3.79 & 5.62 & 3.82 & 8.66 & 2.79 & 12.41\\
1300 & 1000 & 3.88 & 5.74 & 3.90 & 8.85 & 2.85 & 12.69\\
1300 & 1100 & 4.15 & 6.14 & 4.18 & 9.47 & 3.05 & 13.58\\
1300 & 1200 & 4.91 & 7.27 & 4.94 & 11.21 & 3.61 & 16.06\\ \hline
1400 & 100 & 54.25 & 87.60 & 58.80 & 134.53 & 41.56 & 195.65\\
1400 & 200 & 9.42 & 15.20 & 10.20 & 23.35 & 7.21 & 33.95\\
1400 & 300 & 5.10 & 8.23 & 5.52 & 12.64 & 3.90 & 18.38\\
1400 & 400 & 3.93 & 6.35 & 4.26 & 9.75 & 3.01 & 14.18\\
1400 & 500 & 3.46 & 5.58 & 3.75 & 8.57 & 2.65 & 12.46\\
1400 & 600 & 3.18 & 5.14 & 3.45 & 7.89 & 2.44 & 11.48\\
1400 & 700 & 3.07 & 4.95 & 3.33 & 7.61 & 2.35 & 11.06\\
1400 & 800 & 3.00 & 4.85 & 3.25 & 7.44 & 2.30 & 10.83\\
1400 & 900 & 2.98 & 4.80 & 3.22 & 7.38 & 2.28 & 10.73\\
1400 & 1000 & 2.99 & 4.83 & 3.24 & 7.42 & 2.29 & 10.79\\
1400 & 1100 & 3.09 & 4.98 & 3.35 & 7.66 & 2.36 & 11.13\\
1400 & 1200 & 3.25 & 5.25 & 3.52 & 8.06 & 2.49 & 11.72\\
1400 & 1300 & 3.82 & 6.16 & 4.14 & 9.47 & 2.92 & 13.77\\
\hline
\end{tabular}
\label{tab:limits1200to1400}
\end{table}

\begin{table}[!h]
\centering
\topcaption{The 95\% confidence level observed (Obs.) 
and expected (Exp.) exclusion limits (in fb) on the \WR 
production cross section times branching fraction for 
$\WR \rightarrow \mu \mu j j$ as a function of \WR and \Nmu 
mass (in~\GeV) for $1500~\GeV \leq M_{\WR} \leq 1700~\GeV$.  
The 68\% and 95\% uncertainty bands for the expected limit 
(Exp. $\pm 1 \sigma$ and Exp. $\pm 2 \sigma$, respectively) 
are also included for each $(M_{\WR},M_{\Nmu})$ entry.} 
\begin{tabular}{cccccccc}
\hline
$M_{\WR}$   &  $M_{\Nmu}$  & Obs. Limit & Exp. Limit & Exp. $-1 \sigma$  & Exp. $+1 \sigma$ & Exp. $-2 \sigma$  & Exp. $+2 \sigma$  \\
\hline
1500 & 100 & 62.08 & 86.59 & 58.31 & 132.45 & 42.98 & 188.68\\
1500 & 200 & 10.60 & 14.78 & 9.96 & 22.62 & 7.34 & 32.22\\
1500 & 300 & 5.26 & 7.34 & 4.94 & 11.23 & 3.64 & 16.00\\
1500 & 400 & 4.00 & 5.58 & 3.76 & 8.54 & 2.77 & 12.16\\
1500 & 500 & 3.50 & 4.89 & 3.29 & 7.48 & 2.43 & 10.65\\
1500 & 600 & 3.22 & 4.48 & 3.02 & 6.86 & 2.23 & 9.77\\
1500 & 700 & 3.02 & 4.21 & 2.83 & 6.44 & 2.09 & 9.17\\
1500 & 800 & 2.95 & 4.12 & 2.77 & 6.29 & 2.04 & 8.97\\
1500 & 900 & 2.93 & 4.09 & 2.75 & 6.25 & 2.03 & 8.90\\
1500 & 1000 & 2.94 & 4.10 & 2.76 & 6.27 & 2.03 & 8.93\\
1500 & 1100 & 2.92 & 4.08 & 2.75 & 6.24 & 2.02 & 8.89\\
1500 & 1200 & 3.01 & 4.20 & 2.83 & 6.42 & 2.08 & 9.15\\
1500 & 1300 & 3.21 & 4.48 & 3.01 & 6.85 & 2.22 & 9.75\\
1500 & 1400 & 3.71 & 5.17 & 3.48 & 7.91 & 2.57 & 11.27\\ \hline
1600 & 100 & 77.66 & 80.27 & 54.11 & 121.06 & 39.92 & 169.10\\
1600 & 200 & 13.31 & 13.75 & 9.27 & 20.74 & 6.84 & 28.97\\
1600 & 300 & 6.35 & 6.57 & 4.43 & 9.91 & 3.27 & 13.84\\
1600 & 400 & 4.77 & 4.93 & 3.33 & 7.44 & 2.45 & 10.39\\
1600 & 500 & 4.10 & 4.24 & 2.86 & 6.39 & 2.11 & 8.93\\
1600 & 600 & 3.72 & 3.84 & 2.59 & 5.79 & 1.91 & 8.09\\
1600 & 700 & 3.52 & 3.64 & 2.45 & 5.49 & 1.81 & 7.66\\
1600 & 800 & 3.43 & 3.55 & 2.39 & 5.35 & 1.76 & 7.47\\
1600 & 900 & 3.35 & 3.46 & 2.33 & 5.22 & 1.72 & 7.29\\
1600 & 1000 & 3.33 & 3.44 & 2.32 & 5.19 & 1.71 & 7.25\\
1600 & 1100 & 3.35 & 3.47 & 2.34 & 5.23 & 1.72 & 7.30\\
1600 & 1200 & 3.37 & 3.49 & 2.35 & 5.26 & 1.73 & 7.35\\
1600 & 1300 & 3.47 & 3.59 & 2.42 & 5.42 & 1.79 & 7.57\\
1600 & 1400 & 3.66 & 3.78 & 2.55 & 5.70 & 1.88 & 7.96\\
1600 & 1500 & 4.21 & 4.35 & 2.93 & 6.56 & 2.16 & 9.16\\ \hline
1700 & 100 & 98.67 & 79.48 & 53.20 & 119.50 & 38.65 & 176.76\\
1700 & 200 & 16.24 & 13.08 & 8.75 & 19.66 & 6.36 & 29.09\\
1700 & 300 & 7.66 & 6.17 & 4.13 & 9.28 & 3.00 & 13.73\\
1700 & 400 & 5.48 & 4.42 & 2.96 & 6.64 & 2.15 & 9.82\\
1700 & 500 & 4.66 & 3.75 & 2.51 & 5.64 & 1.83 & 8.35\\
1700 & 600 & 4.26 & 3.43 & 2.29 & 5.16 & 1.67 & 7.62\\
1700 & 700 & 4.00 & 3.22 & 2.16 & 4.85 & 1.57 & 7.17\\
1700 & 800 & 3.85 & 3.10 & 2.08 & 4.67 & 1.51 & 6.91\\
1700 & 900 & 3.75 & 3.02 & 2.02 & 4.54 & 1.47 & 6.72\\
1700 & 1000 & 3.73 & 3.01 & 2.01 & 4.52 & 1.46 & 6.69\\
1700 & 1100 & 3.72 & 3.00 & 2.01 & 4.51 & 1.46 & 6.67\\
1700 & 1200 & 3.75 & 3.02 & 2.02 & 4.54 & 1.47 & 6.72\\
1700 & 1300 & 3.76 & 3.03 & 2.03 & 4.56 & 1.47 & 6.74\\
1700 & 1400 & 3.91 & 3.15 & 2.11 & 4.73 & 1.53 & 7.00\\
1700 & 1500 & 4.09 & 3.29 & 2.21 & 4.95 & 1.60 & 7.33\\
1700 & 1600 & 4.64 & 3.74 & 2.50 & 5.62 & 1.82 & 8.31\\
\hline
\end{tabular}
\label{tab:limits1500to1700}
\end{table}

\begin{table}[!h]
\centering
\topcaption{The 95\% confidence level observed (Obs.) 
and expected (Exp.) exclusion limits (in fb) on the \WR 
production cross section times branching fraction for 
$\WR \rightarrow \mu \mu j j$ as a function of \WR and \Nmu 
mass (in~\GeV) for $1800~\GeV \leq M_{\WR} \leq 1900~\GeV$.  
The 68\% and 95\% uncertainty bands for the expected limit 
(Exp. $\pm 1 \sigma$ and Exp. $\pm 2 \sigma$, respectively) 
are also included for each $(M_{\WR},M_{\Nmu})$ entry.} 
\begin{tabular}{cccccccc}
\hline
$M_{\WR}$   &  $M_{\Nmu}$  & Obs. Limit & Exp. Limit & Exp. $-1 \sigma$  & Exp. $+1 \sigma$ & Exp. $-2 \sigma$  & Exp. $+2 \sigma$  \\
\hline
1800 & 100 & 111.90 & 79.36 & 53.08 & 120.16 & 38.73 & 178.56\\
1800 & 200 & 17.80 & 12.62 & 8.44 & 19.12 & 6.16 & 28.41\\
1800 & 300 & 8.40 & 5.96 & 3.99 & 9.03 & 2.91 & 13.41\\
1800 & 400 & 5.84 & 4.14 & 2.77 & 6.27 & 2.02 & 9.32\\
1800 & 500 & 4.91 & 3.48 & 2.33 & 5.27 & 1.70 & 7.84\\
1800 & 600 & 4.44 & 3.15 & 2.11 & 4.77 & 1.54 & 7.09\\
1800 & 700 & 4.13 & 2.93 & 1.96 & 4.43 & 1.43 & 6.58\\
1800 & 800 & 4.01 & 2.84 & 1.90 & 4.30 & 1.39 & 6.40\\
1800 & 900 & 3.89 & 2.76 & 1.84 & 4.17 & 1.34 & 6.20\\
1800 & 1000 & 3.85 & 2.73 & 1.83 & 4.13 & 1.33 & 6.14\\
1800 & 1100 & 3.82 & 2.71 & 1.81 & 4.11 & 1.32 & 6.10\\
1800 & 1200 & 3.82 & 2.71 & 1.81 & 4.11 & 1.32 & 6.10\\
1800 & 1300 & 3.83 & 2.72 & 1.82 & 4.11 & 1.33 & 6.11\\
1800 & 1400 & 3.90 & 2.76 & 1.85 & 4.18 & 1.35 & 6.22\\
1800 & 1500 & 4.00 & 2.83 & 1.90 & 4.29 & 1.38 & 6.38\\
1800 & 1600 & 4.21 & 2.99 & 2.00 & 4.53 & 1.46 & 6.73\\
1800 & 1700 & 4.70 & 3.33 & 2.23 & 5.04 & 1.63 & 7.49\\ \hline
1900 & 100 & 106.11 & 73.41 & 50.20 & 113.01 & 37.43 & 169.08\\
1900 & 200 & 17.65 & 12.21 & 8.35 & 18.80 & 6.23 & 28.13\\
1900 & 300 & 8.14 & 5.63 & 3.85 & 8.67 & 2.87 & 12.98\\
1900 & 400 & 5.55 & 3.84 & 2.63 & 5.91 & 1.96 & 8.85\\
1900 & 500 & 4.54 & 3.14 & 2.15 & 4.84 & 1.60 & 7.24\\
1900 & 600 & 4.12 & 2.85 & 1.95 & 4.39 & 1.45 & 6.57\\
1900 & 700 & 3.85 & 2.66 & 1.82 & 4.10 & 1.36 & 6.13\\
1900 & 800 & 3.66 & 2.53 & 1.73 & 3.90 & 1.29 & 5.83\\
1900 & 900 & 3.57 & 2.47 & 1.69 & 3.80 & 1.26 & 5.68\\
1900 & 1000 & 3.53 & 2.44 & 1.67 & 3.76 & 1.25 & 5.63\\
1900 & 1100 & 3.48 & 2.41 & 1.64 & 3.70 & 1.23 & 5.54\\
1900 & 1200 & 3.46 & 2.39 & 1.64 & 3.68 & 1.22 & 5.51\\
1900 & 1300 & 3.46 & 2.39 & 1.64 & 3.69 & 1.22 & 5.51\\
1900 & 1400 & 3.48 & 2.41 & 1.65 & 3.71 & 1.23 & 5.55\\
1900 & 1500 & 3.53 & 2.44 & 1.67 & 3.76 & 1.24 & 5.62\\
1900 & 1600 & 3.61 & 2.50 & 1.71 & 3.84 & 1.27 & 5.75\\
1900 & 1700 & 3.84 & 2.65 & 1.81 & 4.08 & 1.35 & 6.11\\
1900 & 1800 & 4.25 & 2.94 & 2.01 & 4.53 & 1.50 & 6.77\\
\hline
\end{tabular}
\label{tab:limits1800to1900}
\end{table}

\begin{table}[!h]
\centering
\topcaption{The 95\% confidence level observed (Obs.) 
and expected (Exp.) exclusion limits (in fb) on the \WR 
production cross section times branching fraction for 
$\WR \rightarrow \mu \mu j j$ as a function of \WR and \Nmu 
mass (in~\GeV) for $2000~\GeV \leq M_{\WR} \leq 2100~\GeV$.  
The 68\% and 95\% uncertainty bands for the expected limit 
(Exp. $\pm 1 \sigma$ and Exp. $\pm 2 \sigma$, respectively) 
are also included for each $(M_{\WR},M_{\Nmu})$ entry.} 
\begin{tabular}{cccccccc}
\hline
$M_{\WR}$   &  $M_{\Nmu}$  & Obs. Limit & Exp. Limit & Exp. $-1 \sigma$  & Exp. $+1 \sigma$ & Exp. $-2 \sigma$  & Exp. $+2 \sigma$  \\
\hline
2000 & 100 & 98.21 & 70.58 & 49.13 & 107.35 & 37.07 & 164.39\\
2000 & 200 & 16.04 & 11.53 & 8.02 & 17.53 & 6.05 & 26.85\\
2000 & 300 & 7.32 & 5.26 & 3.66 & 8.00 & 2.76 & 12.26\\
2000 & 400 & 4.92 & 3.53 & 2.46 & 5.37 & 1.86 & 8.23\\
2000 & 500 & 3.99 & 2.86 & 1.99 & 4.36 & 1.50 & 6.67\\
2000 & 600 & 3.56 & 2.56 & 1.78 & 3.90 & 1.35 & 5.96\\
2000 & 700 & 3.31 & 2.38 & 1.65 & 3.61 & 1.25 & 5.53\\
2000 & 800 & 3.16 & 2.27 & 1.58 & 3.46 & 1.19 & 5.30\\
2000 & 900 & 3.08 & 2.21 & 1.54 & 3.37 & 1.16 & 5.16\\
2000 & 1000 & 3.02 & 2.17 & 1.51 & 3.30 & 1.14 & 5.05\\
2000 & 1100 & 2.97 & 2.14 & 1.49 & 3.25 & 1.12 & 4.98\\
2000 & 1200 & 2.96 & 2.13 & 1.48 & 3.24 & 1.12 & 4.96\\
2000 & 1300 & 2.95 & 2.12 & 1.47 & 3.22 & 1.11 & 4.93\\
2000 & 1400 & 2.94 & 2.12 & 1.47 & 3.22 & 1.11 & 4.93\\
2000 & 1500 & 2.99 & 2.15 & 1.50 & 3.27 & 1.13 & 5.01\\
2000 & 1600 & 3.02 & 2.17 & 1.51 & 3.30 & 1.14 & 5.06\\
2000 & 1700 & 3.11 & 2.24 & 1.56 & 3.40 & 1.17 & 5.21\\
2000 & 1800 & 3.28 & 2.36 & 1.64 & 3.58 & 1.24 & 5.49\\
2000 & 1900 & 3.64 & 2.61 & 1.82 & 3.98 & 1.37 & 6.09\\ \hline
2100 & 100 & 82.38 & 64.00 & 45.12 & 96.25 & 36.23 & 151.27\\
2100 & 200 & 13.88 & 10.78 & 7.60 & 16.21 & 6.10 & 25.48\\
2100 & 300 & 6.33 & 4.92 & 3.47 & 7.40 & 2.79 & 11.63\\
2100 & 400 & 4.26 & 3.31 & 2.33 & 4.98 & 1.87 & 7.82\\
2100 & 500 & 3.43 & 2.67 & 1.88 & 4.01 & 1.51 & 6.30\\
2100 & 600 & 3.00 & 2.33 & 1.64 & 3.51 & 1.32 & 5.51\\
2100 & 700 & 2.78 & 2.16 & 1.52 & 3.24 & 1.22 & 5.10\\
2100 & 800 & 2.67 & 2.07 & 1.46 & 3.12 & 1.17 & 4.90\\
2100 & 900 & 2.57 & 1.99 & 1.41 & 3.00 & 1.13 & 4.72\\
2100 & 1000 & 2.50 & 1.94 & 1.37 & 2.92 & 1.10 & 4.59\\
2100 & 1100 & 2.46 & 1.91 & 1.35 & 2.88 & 1.08 & 4.52\\
2100 & 1200 & 2.45 & 1.90 & 1.34 & 2.86 & 1.08 & 4.50\\
2100 & 1300 & 2.45 & 1.91 & 1.34 & 2.87 & 1.08 & 4.51\\
2100 & 1400 & 2.45 & 1.91 & 1.34 & 2.87 & 1.08 & 4.50\\
2100 & 1500 & 2.45 & 1.91 & 1.34 & 2.87 & 1.08 & 4.51\\
2100 & 1600 & 2.47 & 1.92 & 1.35 & 2.89 & 1.09 & 4.53\\
2100 & 1700 & 2.52 & 1.96 & 1.38 & 2.94 & 1.11 & 4.62\\
2100 & 1800 & 2.56 & 1.99 & 1.40 & 3.00 & 1.13 & 4.71\\
2100 & 1900 & 2.68 & 2.09 & 1.47 & 3.14 & 1.18 & 4.93\\
2100 & 2000 & 2.96 & 2.30 & 1.62 & 3.46 & 1.30 & 5.44\\
\hline
\end{tabular}
\label{tab:limits2000to2100}
\end{table}

\begin{table}[!h]
\centering
\topcaption{The 95\% confidence level observed (Obs.) 
and expected (Exp.) exclusion limits (in fb) on the \WR 
production cross section times branching fraction for 
$\WR \rightarrow \mu \mu j j$ as a function of \WR and \Nmu 
mass (in~\GeV) for $2200~\GeV \leq M_{\WR} \leq 2300~\GeV$.  
The 68\% and 95\% uncertainty bands for the expected limit 
(Exp. $\pm 1 \sigma$ and Exp. $\pm 2 \sigma$, respectively) 
are also included for each $(M_{\WR},M_{\Nmu})$ entry.} 
\begin{tabular}{cccccccc}
\hline
$M_{\WR}$   &  $M_{\Nmu}$  & Obs. Limit & Exp. Limit & Exp. $-1 \sigma$  & Exp. $+1 \sigma$ & Exp. $-2 \sigma$  & Exp. $+2 \sigma$  \\
\hline
2200 & 100 & 61.55 & 53.90 & 37.83 & 81.66 & 31.30 & 121.24\\
2200 & 200 & 11.83 & 10.36 & 7.27 & 15.69 & 6.01 & 23.30\\
2200 & 300 & 5.58 & 4.89 & 3.43 & 7.41 & 2.84 & 11.00\\
2200 & 400 & 3.58 & 3.13 & 2.20 & 4.74 & 1.82 & 7.04\\
2200 & 500 & 2.89 & 2.53 & 1.77 & 3.83 & 1.47 & 5.68\\
2200 & 600 & 2.51 & 2.20 & 1.54 & 3.33 & 1.28 & 4.94\\
2200 & 700 & 2.32 & 2.03 & 1.42 & 3.07 & 1.18 & 4.56\\
2200 & 800 & 2.17 & 1.90 & 1.34 & 2.88 & 1.10 & 4.28\\
2200 & 900 & 2.12 & 1.86 & 1.30 & 2.82 & 1.08 & 4.18\\
2200 & 1000 & 2.07 & 1.81 & 1.27 & 2.74 & 1.05 & 4.07\\
2200 & 1100 & 2.03 & 1.78 & 1.25 & 2.69 & 1.03 & 4.00\\
2200 & 1200 & 2.03 & 1.78 & 1.25 & 2.70 & 1.03 & 4.00\\
2200 & 1300 & 2.00 & 1.75 & 1.23 & 2.66 & 1.02 & 3.95\\
2200 & 1400 & 2.00 & 1.75 & 1.23 & 2.66 & 1.02 & 3.94\\
2200 & 1500 & 2.00 & 1.75 & 1.23 & 2.65 & 1.02 & 3.94\\
2200 & 1600 & 2.00 & 1.75 & 1.23 & 2.66 & 1.02 & 3.94\\
2200 & 1700 & 2.02 & 1.77 & 1.24 & 2.68 & 1.03 & 3.99\\
2200 & 1800 & 2.06 & 1.81 & 1.27 & 2.73 & 1.05 & 4.06\\
2200 & 1900 & 2.10 & 1.84 & 1.29 & 2.79 & 1.07 & 4.14\\
2200 & 2000 & 2.20 & 1.93 & 1.35 & 2.92 & 1.12 & 4.34\\
2200 & 2100 & 2.40 & 2.11 & 1.48 & 3.19 & 1.22 & 4.74\\ \hline
2300 & 100 & 53.75 & 52.32 & 37.28 & 79.45 & 31.13 & 115.64\\
2300 & 200 & 10.63 & 10.34 & 7.37 & 15.71 & 6.15 & 22.86\\
2300 & 300 & 4.68 & 4.55 & 3.24 & 6.91 & 2.71 & 10.06\\
2300 & 400 & 3.18 & 3.09 & 2.20 & 4.69 & 1.84 & 6.83\\
2300 & 500 & 2.45 & 2.38 & 1.70 & 3.61 & 1.42 & 5.26\\
2300 & 600 & 2.15 & 2.09 & 1.49 & 3.18 & 1.25 & 4.63\\
2300 & 700 & 1.97 & 1.92 & 1.37 & 2.91 & 1.14 & 4.24\\
2300 & 800 & 1.84 & 1.79 & 1.28 & 2.72 & 1.07 & 3.96\\
2300 & 900 & 1.78 & 1.73 & 1.23 & 2.62 & 1.03 & 3.82\\
2300 & 1000 & 1.73 & 1.69 & 1.20 & 2.56 & 1.00 & 3.73\\
2300 & 1100 & 1.71 & 1.66 & 1.18 & 2.53 & 0.99 & 3.68\\
2300 & 1200 & 1.69 & 1.64 & 1.17 & 2.49 & 0.98 & 3.63\\
2300 & 1300 & 1.68 & 1.63 & 1.16 & 2.48 & 0.97 & 3.61\\
2300 & 1400 & 1.67 & 1.62 & 1.16 & 2.46 & 0.97 & 3.59\\
2300 & 1500 & 1.67 & 1.63 & 1.16 & 2.47 & 0.97 & 3.60\\
2300 & 1600 & 1.66 & 1.62 & 1.15 & 2.46 & 0.96 & 3.58\\
2300 & 1700 & 1.69 & 1.64 & 1.17 & 2.49 & 0.98 & 3.63\\
2300 & 1800 & 1.69 & 1.65 & 1.18 & 2.51 & 0.98 & 3.65\\
2300 & 1900 & 1.72 & 1.67 & 1.19 & 2.54 & 1.00 & 3.70\\
2300 & 2000 & 1.75 & 1.71 & 1.22 & 2.59 & 1.02 & 3.77\\
2300 & 2100 & 1.84 & 1.79 & 1.27 & 2.72 & 1.06 & 3.95\\
2300 & 2200 & 2.02 & 1.96 & 1.40 & 2.98 & 1.17 & 4.34\\
\hline
\end{tabular}
\label{tab:limits2200to2300}
\end{table}

\begin{table}[!h]
\centering
\topcaption{The 95\% confidence level observed (Obs.) 
and expected (Exp.) exclusion limits (in fb) on the \WR 
production cross section times branching fraction for 
$\WR \rightarrow \mu \mu j j$ as a function of \WR and \Nmu 
mass (in~\GeV) for $2400~\GeV \leq M_{\WR} \leq 2500~\GeV$.
The 68\% and 95\% uncertainty bands for the expected limit 
(Exp. $\pm 1 \sigma$ and Exp. $\pm 2 \sigma$, respectively) 
are also included for each $(M_{\WR},M_{\Nmu})$ entry.} 
\begin{tabular}{cccccccc}
\hline
$M_{\WR}$   &  $M_{\Nmu}$  & Obs. Limit & Exp. Limit & Exp. $-1 \sigma$  & Exp. $+1 \sigma$ & Exp. $-2 \sigma$  & Exp. $+2 \sigma$  \\
\hline
2400 & 100 & 43.94 & 43.82 & 32.64 & 66.42 & 28.01 & 100.50\\
2400 & 200 & 9.52 & 9.49 & 7.07 & 14.38 & 6.07 & 21.76\\
2400 & 300 & 4.59 & 4.58 & 3.41 & 6.94 & 2.93 & 10.50\\
2400 & 400 & 2.95 & 2.94 & 2.19 & 4.45 & 1.88 & 6.74\\
2400 & 500 & 2.34 & 2.33 & 1.74 & 3.54 & 1.49 & 5.35\\
2400 & 600 & 2.00 & 1.99 & 1.49 & 3.02 & 1.27 & 4.57\\
2400 & 700 & 1.83 & 1.82 & 1.36 & 2.77 & 1.17 & 4.18\\
2400 & 800 & 1.72 & 1.72 & 1.28 & 2.61 & 1.10 & 3.94\\
2400 & 900 & 1.65 & 1.65 & 1.23 & 2.50 & 1.05 & 3.78\\
2400 & 1000 & 1.62 & 1.62 & 1.21 & 2.45 & 1.03 & 3.71\\
2400 & 1100 & 1.59 & 1.58 & 1.18 & 2.40 & 1.01 & 3.63\\
2400 & 1200 & 1.56 & 1.56 & 1.16 & 2.36 & 1.00 & 3.57\\
2400 & 1300 & 1.54 & 1.54 & 1.15 & 2.33 & 0.98 & 3.53\\
2400 & 1400 & 1.54 & 1.53 & 1.14 & 2.32 & 0.98 & 3.51\\
2400 & 1500 & 1.53 & 1.53 & 1.14 & 2.31 & 0.98 & 3.50\\
2400 & 1600 & 1.52 & 1.52 & 1.13 & 2.30 & 0.97 & 3.49\\
2400 & 1700 & 1.54 & 1.53 & 1.14 & 2.33 & 0.98 & 3.52\\
2400 & 1800 & 1.56 & 1.55 & 1.16 & 2.35 & 0.99 & 3.56\\
2400 & 1900 & 1.57 & 1.56 & 1.16 & 2.37 & 1.00 & 3.59\\
2400 & 2000 & 1.59 & 1.58 & 1.18 & 2.40 & 1.01 & 3.63\\
2400 & 2100 & 1.62 & 1.62 & 1.21 & 2.45 & 1.04 & 3.71\\
2400 & 2200 & 1.69 & 1.69 & 1.26 & 2.56 & 1.08 & 3.87\\
2400 & 2300 & 1.83 & 1.83 & 1.36 & 2.77 & 1.17 & 4.19\\ \hline
2500 & 100 & 42.22 & 44.33 & 34.29 & 67.02 & 29.64 & 102.78\\
2500 & 200 & 9.04 & 9.49 & 7.34 & 14.35 & 6.35 & 22.01\\
2500 & 300 & 4.56 & 4.78 & 3.70 & 7.23 & 3.20 & 11.09\\
2500 & 400 & 2.98 & 3.13 & 2.42 & 4.73 & 2.09 & 7.25\\
2500 & 500 & 2.31 & 2.43 & 1.88 & 3.67 & 1.62 & 5.63\\
2500 & 600 & 1.97 & 2.07 & 1.60 & 3.12 & 1.38 & 4.79\\
2500 & 700 & 1.79 & 1.88 & 1.45 & 2.84 & 1.26 & 4.36\\
2500 & 800 & 1.67 & 1.76 & 1.36 & 2.66 & 1.18 & 4.08\\
2500 & 900 & 1.62 & 1.70 & 1.31 & 2.57 & 1.14 & 3.94\\
2500 & 1000 & 1.58 & 1.66 & 1.28 & 2.51 & 1.11 & 3.85\\
2500 & 1100 & 1.54 & 1.61 & 1.25 & 2.44 & 1.08 & 3.74\\
2500 & 1200 & 1.51 & 1.58 & 1.22 & 2.39 & 1.06 & 3.67\\
2500 & 1300 & 1.50 & 1.57 & 1.22 & 2.38 & 1.05 & 3.65\\
2500 & 1400 & 1.48 & 1.55 & 1.20 & 2.35 & 1.04 & 3.60\\
2500 & 1500 & 1.48 & 1.56 & 1.20 & 2.35 & 1.04 & 3.61\\
2500 & 1600 & 1.48 & 1.55 & 1.20 & 2.34 & 1.04 & 3.59\\
2500 & 1700 & 1.48 & 1.56 & 1.20 & 2.35 & 1.04 & 3.61\\
2500 & 1800 & 1.48 & 1.55 & 1.20 & 2.35 & 1.04 & 3.61\\
2500 & 1900 & 1.49 & 1.56 & 1.21 & 2.36 & 1.04 & 3.62\\
2500 & 2000 & 1.50 & 1.58 & 1.22 & 2.38 & 1.05 & 3.66\\
2500 & 2100 & 1.52 & 1.60 & 1.24 & 2.42 & 1.07 & 3.71\\
2500 & 2200 & 1.56 & 1.64 & 1.27 & 2.48 & 1.10 & 3.80\\
2500 & 2300 & 1.62 & 1.70 & 1.32 & 2.57 & 1.14 & 3.94\\
2500 & 2400 & 1.77 & 1.86 & 1.44 & 2.82 & 1.25 & 4.32\\
\hline
\end{tabular}
\label{tab:limits2400to2500}
\end{table}

}
\cleardoublepage \appendix\section{The CMS Collaboration \label{app:collab}}\begin{sloppypar}\hyphenpenalty=5000\widowpenalty=500\clubpenalty=5000\textbf{Yerevan Physics Institute,  Yerevan,  Armenia}\\*[0pt]
S.~Chatrchyan, V.~Khachatryan, A.M.~Sirunyan, A.~Tumasyan
\vskip\cmsinstskip
\textbf{Institut f\"{u}r Hochenergiephysik der OeAW,  Wien,  Austria}\\*[0pt]
W.~Adam, E.~Aguilo, T.~Bergauer, M.~Dragicevic, J.~Er\"{o}, C.~Fabjan\cmsAuthorMark{1}, M.~Friedl, R.~Fr\"{u}hwirth\cmsAuthorMark{1}, V.M.~Ghete, J.~Hammer, N.~H\"{o}rmann, J.~Hrubec, M.~Jeitler\cmsAuthorMark{1}, W.~Kiesenhofer, V.~Kn\"{u}nz, M.~Krammer\cmsAuthorMark{1}, I.~Kr\"{a}tschmer, D.~Liko, I.~Mikulec, M.~Pernicka$^{\textrm{\dag}}$, B.~Rahbaran, C.~Rohringer, H.~Rohringer, R.~Sch\"{o}fbeck, J.~Strauss, A.~Taurok, W.~Waltenberger, C.-E.~Wulz\cmsAuthorMark{1}
% ...
\vskip\cmsinstskip
\dag:~Deceased\\
1:~~Also at Vienna University of Technology, Vienna, Austria\\
% ...

\end{sloppypar}
\end{document}